\documentstyle[aps,prl,epsfig,twocolumn,floats]{revtex}

\def\dh{{\delta_{H}}}
\def\n{{\bf n}}
\def\k{{\bf k}}

\def\beq#1{\begin{equation}\label{#1}}
\def\eeq{\end{equation}}
\def\beqa#1{\begin{eqnarray}\label{#1}}
\def\eeqa{\end{eqnarray}}
\def\eq#1{equation~(\ref{#1})}

\begin{document}

\draft

\tighten

\twocolumn[\hsize\textwidth\columnwidth\hsize\csname@twocolumnfalse\endcsname

\title{Measuring the Small-Scale Power Spectrum of Cosmic Density
Fluctuations \\ Through 21 cm Tomography Prior to the Epoch of Structure
Formation}

\author{Abraham Loeb$^\star$ \& Matias Zaldarriaga$^{\star,\dagger}$}

\address{$\star$ Astronomy Department, Harvard University, 60 Garden
Street, Cambridge, MA 02138\\ $\dagger$ Physics
Department, Harvard University, 17 Oxford Street, Cambridge MA 02138}
\date{\today} \maketitle

\begin{abstract}

The thermal evolution of the cosmic gas decoupled from that of the cosmic
microwave background (CMB) at a redshift $z\sim 200$. Afterwards and before
the first stars had formed, the cosmic neutral hydrogen absorbed the CMB
flux at its resonant 21cm spin-flip transition.  We calculate the evolution
of the spin temperature for this transition and the resulting anisotropies
that are imprinted on the CMB sky due to linear density fluctuations during
this epoch. These anisotropies at an observed wavelength of $10.56 \
\{(1+z)/50\}$ meters, contain an amount of information that is orders of
magnitude larger than any other cosmological probe.  Their detection,
although challenging, could tightly constrain any possible running of the
spectral index from inflation (as suggested by {\it WMAP}), small
deviations from Gaussianity, or any significant contribution from neutrinos
or warm dark matter to the cosmic mass budget.

\end{abstract}
\pacs{PACS numbers: 98.80.-k, 98.65.-r, 98.70.Vc, 95.30.Jx}

%\bigskip
]

\narrowtext

\paragraph*{Introduction.}
The small residual fraction of free electrons after cosmological
recombination coupled the temperature of the cosmic gas to that of the
cosmic microwave background (CMB) down to a redshift, $z\sim 200$
\cite{Peebles}. Subsequently, the gas temperature dropped adiabatically as
$T_{\rm gas}\propto (1+z)^2$ below the CMB temperature $T_{\gamma}\propto
(1+z)$.  The gas heated up again after being exposed to the photo-ionizing
ultraviolet light emitted by the first stars during the {\it reionization
epoch} at $z\lesssim 20$ (see review in \cite{BL01}). Prior to the
formation of the first stars, the cosmic neutral hydrogen must have
resonantly absorbed the CMB flux through its spin-flip 21cm transition
\cite{Field,Scott,Tozzi,Zalda}.  The linear density fluctuations at that time
should have imprinted anisotropies on the CMB sky at an observed wavelength
of $21.12[(1+z)/100]$ meters.  In this {\it Letter}, we calculate the power
spectrum of these anisotropies and assess the significance of their
potential detection.  A direct measurement of the amplitude of
inhomogeneities on small spatial scales would constrain any possible tilt
or running of the spectral index of the power spectrum as recently
suggested by {\it Wilkinson Microwave Anisotropy Probe} ({\it WMAP})
\cite{WMAP}, or any suppression of power on small scales due to a warm dark
matter component in the cosmic mass budget \cite{Ba01}.

\paragraph*{Spin temperature history.} We start by  calculating the 
history of the spin temperature, $T_s$, defined through the ratio between
the number densities of hydrogen atoms in the excited and ground state
levels, ${n_1/ n_0}=(g_1/ g_0)\exp\left\{-{T_\star/ T_s}\right\},$
%\begin{equation}
%{n_1\over n_0}={g_1\over g_0}\exp\left\{-{T_\star\over T_s}\right\},
%\label{eq:spin}
%\end{equation}
where subscripts $1$ and $0$ correspond to the excited and ground state
levels of the 21cm transition, $(g_1/g_0)=3$ is the ratio of the spin
degeneracy factors of the levels, $n_{\rm H}=(n_0+n_1)\propto (1+z)^3$ is
the total hydrogen density, and $T_\star=0.068$K is the temperature
corresponding to the energy difference between the levels.  The time
evolution of the density of atoms in the ground state is given by,
\begin{eqnarray}
\left( \partial_t  +   3{{\dot a}\over a} \right)
& n_0 & =-n_0\left(C_{01}+B_{01}I_\nu\right) \nonumber \\
 + & n_1 & \left(C_{10}+A_{10}+B_{10}I_\nu\right),
\label{eq:evolution}
\end{eqnarray}
where $a(t)=(1+z)^{-1}$ is the cosmic scale factor, $A$'s and $B$'s are the
Einstein rate coefficients, $C$'s are the collisional rate coefficients,
and $I_\nu$ is the blackbody intensity in the Rayleigh-Jeans tail of the
CMB, namely $I_\nu=2k_BT_{\gamma}/\lambda^2$ with $\lambda=21$ cm
\cite{RL}.  The $0\rightarrow 1$ transition rates can be related to the
$1\rightarrow 0$ transition rates by the requirement that in thermal
equilibrium with $T_s=T_\gamma=T_{\rm gas}$, the right-hand-side of
Eq. (\ref{eq:evolution}) should vanish with the collisional terms balancing
each other separately from the radiative terms. The Einstein coefficients
are $A_{10}=2.85\times 10^{-15}~{\rm s^{-1}}$, $B_{10}=(\lambda^3/2hc)
A_{10}$ and $B_{01}=(g_1/g_0)B_{10}$ \cite{Field,RL}.  The collisional
de-excitation rates can be written as $C_{10}={4\over 3} \kappa(1-0) n_{\rm
H}$, where $\kappa(1-0)$ is tabulated as a function of $T_{\rm gas}$
\cite{AD}.

\begin{figure}[th]
\centerline{\epsfig{file=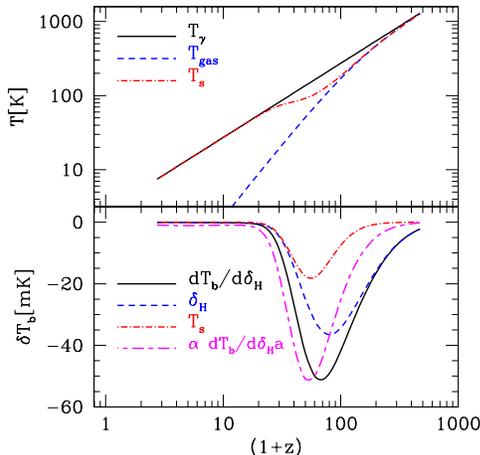,height=2.5in}}
\caption{{\it Upper panel:} Evolution of the gas, CMB and spin temperatures with
redshift [4]. 
{\it Lower panel:} ${dT_b/d\dh}$ as function of redshift. 
The separate contributions from fluctuations in the density and 
the spin temperature are depicted. We also show ${dT_b/d\dh} a
\propto {dT_b/d\dh} \times \dh$, with an arbitrary normalization.
Throughout this {\it Letter}, we assume the standard set of
cosmological parameters for a universe dominated by cold dark matter
and a cosmological constant ($\Lambda$CDM) [6].}
\label{dtb}
\end{figure}

Equation (\ref{eq:evolution}) can
be simplified to the form,
\begin{eqnarray}
{d\Upsilon \over dz} & = & -\left[H(1+z)\right]^{-1}
\left[-\Upsilon(C_{01}+B_{01}I_\nu) \right. \nonumber \\ 
&& \left. +
(1-\Upsilon)(C_{10}+A_{10}  + B_{10}I_\nu)\right],
\label{eq:upsilon}
\end{eqnarray}
where $\Upsilon\equiv n_0/n_{\rm H}$, $H\approx
H_0\sqrt{\Omega_m}(1+z)^{3/2}$ is the Hubble parameter at high redshifts
(with a present-day value of $H_0$), and $\Omega_m$ is the density
parameter of matter.  The upper panel of Fig. \ref{dtb} shows the results
of integrating Eq.~(\ref{eq:upsilon}). Both the spin temperature and the
kinetic temperature of the gas track the CMB temperature down to $z\sim
200$. Collisions are efficient at coupling $T_s$ and $T_{gas}$ down to
$z\sim 70$ and so the spin temperature follows the kinetic temperature
around that redshift. At much lower redshifts, the Hubble expansion makes
the collision rate subdominant relative the radiative coupling rate to the
CMB, and so $T_s$ tracks $T_{\gamma}$ again. Consequently, there is a
redshift window between $30\lesssim z \lesssim 200$, during which the
cosmic hydrogen absorbs the CMB flux at its resonant 21cm
transition. Coincidentally, this redshift interval precedes the appearance
of collapsed objects \cite{BL01} and so its signatures are not contaminated by
nonlinear density structures or by radiative or hydrodynamic feedback
effects from stars and quasars.
%, as is the case at lower redshifts
%\cite{Zalda}.

During the period when the spin temperature is smaller than the CMB
temperature, neutral hydrogen atoms absorb CMB photons. The resonant 21cm
absorption reduces the brightness temperature of the CMB by,
\begin{equation}
T_b =\tau \left( T_s-T_{\gamma}\right)/(1+z) ,
\end{equation}
where the optical depth for resonant 21cm absorption is,
\begin{equation}
\tau= {3c\lambda^2hA_{10}n_{\rm H}\over 32 \pi k_B T_s H(z)} .
\label{eq:tau} 
\end{equation}

Small inhomogeneities in the hydrogen density $\dh\equiv (n_{\rm H}-{\bar
n_{\rm H}})/{\bar n}_{\rm H}$ result in fluctuations of the 21cm
absorption through two separate effects. An excess of neutral hydrogen 
directly increases the optical depth and also alters the evolution of
the spin temperature. We can write an equation for the resulting evolution of
$\Upsilon$ fluctuations,
\begin{eqnarray}
{d\delta \Upsilon \over dz} & = & \left[H(1+z)\right]^{-1}
\left\{[C_{10}+C_{01}+(B_{01}+B_{10})I_\nu]\delta\Upsilon \right. \nonumber \\
 && \left. + \left[ C_{01} \Upsilon
-C_{10}(1-\Upsilon)\right]\dh\right\},
\label{eq:dup}
\end{eqnarray}
leading to spin temperature fluctuations,
\begin{equation}
{\delta T_s\over {\bar T}_s}= -{1\over
\ln[3\Upsilon/(1-\Upsilon)]}{\delta\Upsilon\over \Upsilon (1-\Upsilon)}.
\label{eq:dts}
\end{equation}
The resulting brightness temperature fluctuations can be related to the derivative,
%\begin{equation}
%{\delta T_b\over {\bar T}_b}=\dh +  {T_{\gamma}\over ({\bar T}_s-T_{\gamma})}
%{\delta T_s\over {\bar T}_s}.
%\label{eq:dtb}
%\end{equation}
%The spin temperature fluctuations ${\delta T_s/ {T}_s}$ are proportional to
%the density fluctuations and so we define, 
\beq{weight} 
{d T_b \over d \dh} \equiv {\bar T}_b + {T_{\gamma} {\bar T}_b
\over ({\bar T}_s-T_{\gamma})} {\delta T_s\over {\bar T}_s \dh}, 
\eeq
through ${\delta T_b}=({d T_b /d \dh}) \dh$.  We include all fluctuations
caused by $\dh$ except for the variation in $C_{ij}$ due to
fluctuations in $T_{\rm gas}$ which is very small \cite{AD}.
Figure \ref{dtb} shows ${dT_b/d\dh}$ as a function of redshift, including
the two contributions to ${dT_b/d\dh}$, one originating directly from
density fluctuations and the second from the associated changes in the spin
temperature \cite{Scott}.  Both contributions have the same sign, because
an increase in density raises the collision rate and lowers the spin
temperature and so it allows $T_s$ to better track $T_{\rm gas}$.  Since
$\dh$ grows with time as $\dh \propto a$, the signal peaks at $z\sim 50$, a
slightly lower redshift than the peak of ${dT_b/d\dh}$.

Next we calculate the angular power spectrum of the brightness temperature
on the sky, resulting from density perturbations with a power spectrum
$P_{\delta}(k)$, \beq{pdel} \langle \dh(\k_1) \dh(\k_2) \rangle = (2\pi)^3
\delta^D(\k_1+\k_2) P_{\delta}(k_1).  \eeq where $\dh(\k)$ is the Fourier
tansform of the hydrogen density field, $\k$ is the comoving wavevector,
and $\langle \cdots \rangle$ denotes an ensemble average (following the
formalism described in \cite{Zalda}).  The 21cm brightness temperature
observed at a frequency $\nu$ corresponding to a distance $r$ along the
line of sight, is given by \beq{los} \delta T_b(\n,\nu)=\int dr W_{\nu}(r)
\ {dT_b \over d\dh} \dh(\n,r), \eeq where $\n$ denotes the direction of
observation, $W_{\nu}(r)$ is a narrow function of $r$ that peaks at the
distance corresponding to $\nu$. The details of this function depend on the
characteristics of the experiment.  The brightness fluctuations in \eq{los}
can be expanded in spherical harmonics with expansion coefficients
$a_{lm}(\nu)$. The angular power spectrum of map $C_{l}(\nu)= \langle
|a_{lm}(\nu)|^2 \rangle$ can be expressed in terms of the 3D power spectrum
of fluctuations in the density $P_{\delta}(k)$, \beqa{cldef} C_{l}(\nu)&=&4
\pi \int {d^3k \over (2\pi)^3} P_{\delta}(k) \alpha_l^2(k,\nu) \nonumber \\
\alpha_l(k,\nu)&=& \int dr W_{r_0}(r) {dT_b\over d\dh}(r) j_l(kr).  \eeqa
Our calculation ignores inhomogeneities in the hydrogen ionization
fraction, since they freeze at the earlier recombination epoch ($z\sim
10^3$) and so their amplitude is more than an order of magnitude smaller
than $\dh$ at $z\lesssim 100$.  The peculiar velocity and gravitational
potential perturbations induce redshift distortion effects that are of
order $\sim (H/ck)$ and $\sim (H/ck)^2$ smaller than $\dh$ for the
high--$l$ modes of interest here.  These effects are expected to be
washed-out within realistically broad band filters.

\begin{figure}[th]
\centerline{\epsfig{file=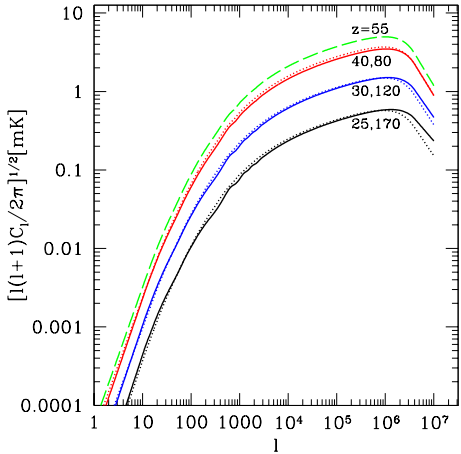,
height=2.5in}}
\caption{Angular power spectrum of 21cm anisotropies on the sky at
various redshifts.  From
top to bottom, $z=55,40,80,30,120,25,170$.}
\label{clfig}
\end{figure}

Figure \ref{clfig} shows the angular power spectrum at various
redshifts. 
%The signal peaks around $z\sim 50$ but maintains a substantial
%amplitude over the full range of $30\lesssim z\lesssim 100$.  
The ability to probe the small scale power of density fluctuations is only
limited by the Jeans scale, below which the dark matter inhomogeneities are
washed out by the finite pressure of the gas. Interestingly, the
cosmological Jeans mass reaches its minimum value, $\sim 3\times 10^4
M_\odot$, within the redshift interval of interest here \cite{BL01}. During
the epoch of reionization, photoionization heating raises the Jeans mass by
several orders of magnitude and and broadens spectral features, thus limiting
the ability of other probes of the intergalactic medium, such as the
Ly$\alpha$ forest, from accessing the same very low mass scales. The 21cm
tomography has the additional advantage of probing the majority of the
cosmic gas, instead of the trace amount ($\sim 10^{-5}$) of neutral
hydrogen probed by the Ly$\alpha$ forest after reionization.  Similarly to
the primary CMB anisotropies, the 21cm signal is simply shaped by gravity,
adiabatic cosmic expansion, and well-known atomic physics, and is not
contaminated by complex astrophysical processes that affect the
intergalactic medium at $z\lesssim 30$.

\paragraph*{The small scale power spectrum.} 
%Characterizing the initial fluctuations is one of the
%primary goals of observational cosmology, as it offers a window into the
%physics of the very early universe, namely the epoch of inflation during
%which the fluctuations are believed to have been produced. 
In most models of inflation, the evolution of the Hubble parameter during
inflation leads to departures from a scale-invariant spectrum that are of
order $1/N_{\rm efold}$ with $N_{\rm efold}\sim 60$ being the number of
$e$--folds between the time when the scale of our horizon was of order the
horizon during inflation and the end of inflation \cite{lidlith}. Recent
{\it WMAP} data combined with other measures of the power on smaller
scales, suggests that the power spectrum changes with scale much faster
than inflation would have predicted \cite{WMAP}, although this result is
still somewhat controversial.  Independent hints that the standard
$\Lambda$CDM model may have too much power on galactic scales have inspired
several proposals for suppressing the power on small scales. Examples
include the possibility that the dark matter is warm and it decoupled while
being relativistic so that its free streaming erased small-scale power
\cite{Ba01}, or direct modifications of inflation that produce a cut-off in
the power on small scales \cite{kamlidd}. An unavoidable collisionless
component of the cosmic mass budget beyond CDM, is provided by massive
neutrinos (see \cite{neut} for a review). Particle physics experiments
established the mass splittings among different species which translate
into a lower limit on the fraction of the dark matter accounted for by
neutrinos of $f_\nu > 0.3 \%$, while current constraints based on galaxies
as tracers of the small scale power imply $f_\nu < 12 \%$ \cite{tegsdss}.

\begin{figure}[th]
\centerline{\epsfig{file=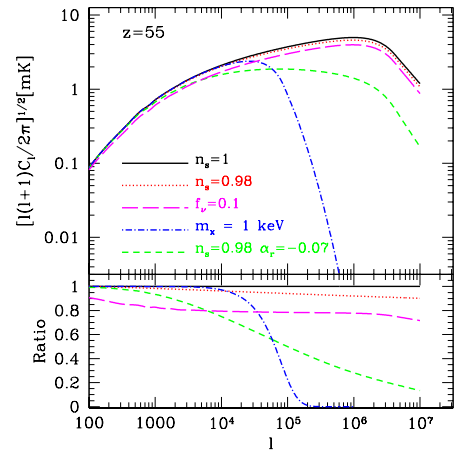,
height=2.5in}}
\caption{{\it Upper panel:} Power spectrum of 21cm anisotropies at $z=55$
for a $\Lambda$CDM scale-invariant power spectrum, a model with $n=0.98$, a
model with $n=0.98$ and $\alpha_r\equiv {1\over 2} (d^2\ln P/d\ln
k^2)=-0.07$, a model of warm dark matter particles with a mass of 1 keV,
and a model in which $f_\nu=10\% $ of the matter density is in three
species of massive neutrinos with a mass of $0.4~{\rm eV}$ each. {\it Lower
panel:} Ratios between the different power spectra and the scale-invariant
spectrum. }
\label{clcomp}
\end{figure}

In Fig. \ref{clcomp} we show the 21cm power spectrum for various
models that differ in their level of small scale power. It is clear that a
precise measurement of the 21cm power spectrum will dramatically improve
current constraints on alternatives to the standard $\Lambda$CDM spectrum.

\paragraph*{Unprecedented information.} 
The 21cm signal contains a wealth of information about the 
initial fluctuations. A full sky map at a single photon 
frequency measured up to $l_{\rm max}$, can 
probe the power spectrum up to $k_{\rm max}\sim (l_{\rm max}/10^4) {\rm
Mpc}^{-1}$. Such a map contains $l_{\rm max}^2$ independent samples. By
shifting the photon frequency, one may obtain many independent measurements
of the power. When measuring a mode $l$, which corresponds to a wavenumber
$k\sim l/r$, two maps at different photon frequencies will be independent
if they are separated in radial distance by $1/k$. Thus, an experiment that
covers a spatial range $\Delta r$ can probe a total of $k\Delta r\sim l
\Delta r/r$ independent maps. An experiment that detects the 21cm signal
over a range $\Delta\nu$ centered on a frequency $\nu$, is sensitive to
$\Delta r/r\sim 0.5 (\Delta\nu/\nu)(1+z)^{-1/2}$, and so it 
measures a total of $N_{\rm 21cm}\sim 3 \times 10^{16} (l_{\rm max}/10^6)^3
(\Delta\nu/\nu) (z/100)^{-1/2}$ independent samples.

This detection capability cannot be reproduced even remotely by other
techniques. For example, the primary CMB anisotropies are damped on small
scales (through the so-called Silk damping), and probe only modes with $l
\leq 3000$ ($k\leq 0.2 \ {\rm Mpc}^{-1}$). The total number of modes
available in the full sky is $N_{\rm cmb} = 2 l_{\rm max}^2\sim 2\times
10^7 (l_{\rm max}/3000)^2$, including both temperature and polarization
information.

\paragraph*{Detectability of signal.}
The sensitivity of an experiment depends strongly on its particular design,
involving the number and distribution of the antennae for an
interferometer. Crudely speaking, the uncertainty in the measurement of
$[{l(l+1)C_l/2\pi}]^{1/2}$ is dominated by noise, $N_\nu$, which is
controlled by 
the sky brightness $I_\nu$ at the observed frequency $\nu$ \cite{Zalda}, 
\beqa{err} 
N_\nu && \sim
0.4 {\rm mK }\left({I_\nu \over 5\times 10^{5} { \rm Jy\ sr^{-1}}}\right)
\left({l_{\rm min}\over 35}\right) \left( {5000 \over
l_{\rm max}}\right) \left( {0.016 \over f_{\rm cover}}\right)
\nonumber \\&& \times \left({1 \ {\rm
year} \over t_0}\right)^{1/2} \left({\Delta \nu \over \nu 
%\over 0.1
}\right)^{-1/2} \ \left({50\ {\rm MHz }\over \nu}\right)^{5/2} , 
\eeqa
where $l_{\rm min}$ is the minimum observable $l$ as determined by the
field of view of the instruments, $l_{\rm max}$ is the maximum observable
$l$ as determined by the maximum separation of the antennae, $f_{\rm
cover}$ is the fraction of the array area thats is covered by telescopes,
$t_0$ is the observation time and $\Delta \nu$ is the frequency range over
which the signal can be detected.
%We have already included the fact that several independent maps 
%can be produced by varying the observed frequency.  
The numbers adopted above are appropriate for the inner core of the {\it
LOFAR} array ({\it http://www.lofar.org}), planned for initial operation in
2006. The predicted signal is $\sim 1{\rm mK}$, and so a year of
integration or an increase in the covering fraction are required to observe
it with {\it LOFAR}.  Other experiments whose goal is to detect 21cm
fluctuations include {\it SKA} ({\it http://www.skatelescope.org}) and {\it
PAST} ({\it http://astrophysics.phys.cmu.edu/}$\sim${\it jbp}).  The main
challenge in detecting the predicted signal involves its appearance at low
frequencies where the sky noise is high.  Proposed space-based instruments
\cite{RadioAstr} avoid the terrestrial radio noise and the increasing
atmospheric opacity at $\nu\lesssim 20 \ {\rm MHz}$ (corresponding to
$z\gtrsim 70$).

\paragraph*{Final comments.}
The 21cm absorption is replaced by 21cm emission from neutral hydrogen as
soon as the intergalactic medium is heated above the CMB temperature during
the epoch of reionization \cite{Miralda}.  Once most of the cosmic hydrogen
is reionized at $z_{\rm reion}$, the 21cm signal is diminished. The optical
depth for free-free absorption after reionization, $\sim 0.1 [(1+z_{\rm
reion})/20]^{5/2}$, modifies only slightly the expected 21cm anisotropies.
Gravitational lensing should modify the power spectrum \cite{Pen} at high
$l$, but can be separated as in standard CMB studies (see \cite{lensing}
and references therein).  The 21cm signal should be simpler to clean as it
includes the same lensing foreground in independent maps obtained at
different frequencies.

The large number of independent modes probed by the 21cm signal would
provide a measure of non-Gaussian deviations to a level of $\sim N_{\rm 21
cm}^{-1/2}$, constituting a test of the inflationary origin of the
primordial inhomogeneities which are expected to possess deviations
$\gtrsim 10^{-6}$ \cite{malda}.
 
\paragraph*{Acknowledgments.}

This work was supported in part by NASA grant NAG 5-13292, NSF grants
AST-0071019, AST-0204514 (for A.L.) and by NSF grants AST-0098606,
PHY-0116590 and the David \& Lucille Packard Foundation Fellowship (for
M.Z.).

\end{document}